\documentclass[prd,aps,nofootinbib]{revtex4} 
\usepackage[sumlimits]{amsmath}
\usepackage{amsmath}
\usepackage{epsfig}
\usepackage{amssymb}
\usepackage{graphicx}
\usepackage{pstricks}
\usepackage{color}
\usepackage{bbold}
\usepackage{hyperref}
\usepackage{subfigure}

\newcommand{\be}{\begin{equation}}
\newcommand{\ee}{\end{equation}}
\newcommand{\bea}{\begin{eqnarray}}
\newcommand{\eea}{\end{eqnarray}}

\begin{document}
\title{Warm inflection}
\author{Rafael Cerezo}
\email{cerezo@ugr.es}
\affiliation{Departamento de F\'{\i}sica Te\'orica y del Cosmos and CAFPE \\  Universidad de Granada, Granada-18071, Spain}

\author{Jo\~ao G. Rosa}
\email{joao.rosa@ua.pt}
\affiliation{Departamento de F\'{\i}sica da Universidade de Aveiro and I3N \\ Campus de Santiago, 3810-183 Aveiro, Portugal}

\begin{abstract}
While ubiquitous in supersymmetric and string theory models, inflationary scenarios near an inflection point in the scalar potential generically require a severe fine-tuning of a priori unrelated supersymmetry breaking effects. We show that this can be significantly alleviated by the inclusion of dissipative effects that damp the inflaton's motion and produce a nearly-thermal radiation bath. We focus on the case where the slow-rolling inflaton directly excites heavy virtual modes that then decay into light degrees of freedom, although our main qualitative results should apply in other regimes. Furthermore, our analysis shows that the minimum amount of dissipation required to keep the temperature of the radiation bath above the Hubble rate during inflation is largely independent of the degree of flatness of the potential, although it depends on the field value at the inflection point. We then discuss the relevance of this result to warm inflation model building.
\end{abstract}

\maketitle


\section{Introduction}

Inflation is an extremely successful paradigm in addressing the shortcomings of standard Big Bang cosmology, providing an elegant solution to the horizon and flatness problems by considering a period of accelerated expansion in the early universe \cite{inflation}. The simplest and most commonly considered scenario is that of a scalar field, the inflaton $\phi$, that slowly rolls down its potential, acting as an effective cosmological constant for a finite period. This typically ends with the field rolling towards the minimum of its potential and, while oscillating about it, reheating the universe through perturbative and non-perturbative particle production. This scenario also provides a seed for the observed temperature anisotropies in the Cosmic Microwave Background and the Large Scale Structure of the universe, as the accelerated expansion amplifies small quantum fluctuations of the scalar field.

The slow-roll inflationary paradigm requires the scalar potential to have a very flat region, where its motion is overdamped and the nearly constant potential is the dominant source of energy density in the early universe. While a plethora of phenomenological potentials with this property have been constructed in the literature, the main challenge has been to embed the inflationary dynamics within a more fundamental theory that reduces to the Standard Model at low energies. This is important not only in establishing a connection between inflation and low-energy particle phenomenology but also due to the sensitivity of the inflationary dynamics to ultraviolet effects close to the Planck scale. 

This has motivated a search for inflaton candidates in supersymmetric (SUSY) theories, in particular in the context of supergravity/string theory (see e.g. \cite{Baumann:2009ni}), which provides the best-known candidate for a fundamental theory of quantum gravity. These scenarios have the appealing feature of naturally including several additional scalar fields, in particular the superpartners of the Standard Model fermions and also the Higgs fields, as well as a plethora of extra-dimensional moduli. Moreover, these models generically exhibit a multitude of directions in field space along which the scalar potential is completely flat in the supersymmetric limit and which are uplifted by different SUSY breaking effects. For example, even the simplest supersymmetric extension of the Standard Model, the MSSM, includes nearly 300 flat directions corresponding to gauge invariant combinations of the matter and Higgs superfields \cite{Gherghetta:1995dv}.

Flat directions can be lifted by a variety of effects, including soft terms from SUSY breaking in a hidden/sequestered sector, renormalizable and non-renormalizable terms in the superpotential, as well as non-perturbative effects (e.g. gaugino condensation). In the context of string theory, these effects are generically related to the geometry and topology of the compactified extra-dimensions, which typically involves different fluxes and/or D-brane configurations. All these different effects may {\it a priori} yield both attractive and repulsive contributions to the scalar potential, which may conspire to produce an inflection point or even a saddle point in the potential. 

The resulting flatness thus provides a very attractive setup for inflationary dynamics, and several successful models have been constructed in the literature. In the context of the MSSM flat directions, inflection points may for example result from the interplay between repulsive soft trilinear $A$-terms and (non-)renormalizable terms in the superpotential, providing not only inflationary models consistent with observational data but also interesting connections to low-energy phenomenology, such as neutrino masses, natural dark matter candidates and the recent Higgs  mass from LHC \cite{Allahverdi:2006iq, Allahverdi:2006cx, Bueno Sanchez:2006xk, Allahverdi:2006we, Allahverdi:2007vy, Allahverdi:2007wt, Lalak:2007rsa, Allahverdi:2010zp, Enqvist:2010vd, Allahverdi:2011su, Chatterjee:2011qr, Boehm:2012rh}. Several different flat directions in the MSSM field space have been analyzed so far, including simple extensions such as additional singlet fields leading to hybrid inflation models \cite{Hotchkiss:2011am}, as well as taking into account supergravity corrections \cite{Enqvist:2007tf, Mazumdar:2011ih} and possible embeddings within the string theory landscape \cite{Allahverdi:2007wh}. In the context of string theory, several new possibilities arise, as for example the case of warped D-brane inflation \cite{Kachru:2003sx, warped_brane_potentials}, where the D-brane potential receives a broad array of contributions such as Coulomb-like interactions in brane-antibrane pairs, couplings to the four-dimensional scalar curvature and several different moduli stabilization effects in the bulk of the compactification. A recent statistical analysis of these contributions has shown that successful models typically occur near an inflection point in the potential \cite{Agarwal:2011wm}. Such features may also be found in closed-string moduli dynamics, for example in the context of racetrack models \cite{BlancoPillado:2004ns} and the so-called accidental inflation scenarios \cite{Linde:2007jn, BlancoPillado:2012cb}.

Inflection point inflation thus appears in a broad range of different  setups, being quite successful in terms of consistency with  observational data, as well as providing a natural embedding within  ultraviolet completions of the Standard Model and desirable links to  low-energy phenomenology. However, these models are far from generic  and typically require a fine-tuning of the different contributions to  the scalar potential, making inflation rather special within the vast  landscape of different possibilities.

In this work, we revisit inflationary dynamics near an inflection point in the potential taking into account the effects of dissipation in the inflaton's motion. Dissipation is a natural outcome of the interactions between the inflaton field and other degrees of freedom, which are in fact required to ensure a graceful exit into a radiation-dominated era, and is intrinsically associated with particle production. If such dissipative effects are sufficiently strong, particle production may actually balance the dilution effect of the accelerated expansion, resulting in an inflationary state that is far from the supercooled vacuum that is conventionally considered. In particular, if the resulting particles have sufficiently strong interactions between them, they can possibly reach a nearly-thermal state at a temperature $T>H$, where $H$ is the inflationary Hubble rate, thus potentially changing the dynamics of inflation. Although it encompasses a much more general class of models where fluctuation-dissipation effects are significant during inflation, the resulting paradigm is thus known as {\it warm inflation}, having been originally proposed in \cite{wi, Berera:1996nv} following on earlier work in \cite{earlydissp}. 

Most scenarios considered in the literature so far in the context of quantum field theory \cite{Berera:1998gx, Yokoyama:1998ju, Berera:1999ws, BR1, Berera:2002sp, Hall:2004zr, br05, Moss:2006gt, Berera:2008ar, Moss:2008yb, Graham:2008vu, BasteroGil:2009ec, BasteroGil:2010pb, BasteroGil:2012cm}, as well as phenomenological models \cite{ph} consider the case where a nearly-thermal bath of radiation is produced concurrently with accelerated expansion from the adiabatic motion of the inflaton field, which may be analyzed within the framework of linear response theory. In this regime, the non-local effects of dissipation yield, to leading order, an additional friction term in the inflaton's equation of motion that helps overdamping its trajectory, thus allowing for longer periods of slow-roll inflation and alleviating the need for a very flat potential (see e.g. \cite{BasteroGil:2009ec}), which is particularly important in the context of supergravity and string theory \cite{warm_string, warm_brane}, where one typically finds a severe `eta-problem'. In this sense, we expect the inclusion of dissipative effects to minimize the fine-tuning of different terms in the scalar potential required for a sufficiently long period of inflation near an inflection point. On the other hand, most studies of warm inflation so far have focused on relatively steep potentials, typically suffering from an eta-problem, where dissipative effects may play a more prominent role, so that this will allow us to explore a new regime of warm inflationary dynamics with a very flat potential.

Warm inflation has several other attractive features, such as providing an alternative to the graceful exit problem, since radiation, although necessarily sub-leading during inflation, may become the dominant component of the energy density as soon as the conditions for slow-roll evolution break down. The spectrum of primordial perturbations is also typically modified in this context, as thermal fluctuations of the inflaton field overcome vacuum fluctuations for $T>H$ \cite{wi, Berera:1999ws,Taylor:2000ze, Hall:2003zp, Moss:2008yb}. This generically suppresses the amount of primordial gravity waves produced during inflation and induces a significant non-gaussian component in the spectrum, which is generically expected to be within the reach of the Planck mission \cite{Gupta:2002kn, Chen:2007gd, Moss:2007cv, Moss:2011qc}. Furthermore, the inclusion of dissipative effects and the finite temperature during inflation may address other outstanding problems in modern cosmology, such as the generation of a baryon asymmetry \cite{BasteroGil:2011cx} and the overproduction of gravitinos in supersymmetric models \cite{Taylor:2000jw, Sanchez:2010vj, Bartrum:2012tg}.

This work is organized as follows. In the next section, we review the conventional dynamics of inflation near an inflection point in the potential, taking as a working example that we will follow throughout our discussion a renormalizable flat direction in a $U(1)_{\mathrm{B-L}}$ extension of the MSSM. In section III we give a basic review of the generic features and conditions for a successful model of warm inflation, applying these to a potential with an inflection point in section IV. We then perform numerical simulations of the evolution of the coupled inflaton-radiation system as a function of the parameters characterizing the potential and the dissipation coefficient, presenting our results in section V. Finally, in section VI we summarize our main conclusions and discuss their impact on inflationary model building, as well as prospects for future research in this topic.


\section{Cold inflation near an inflection point}

As discussed earlier, scalar potentials exhibiting an inflection point may arise in a variety of models in supersymmetric theories and supergravity/string theory models. For concreteness, we will consider throughout most of our discussion a simple example introduced in \cite{Allahverdi:2006iq, Allahverdi:2007vy} and also considered in \cite{Hotchkiss:2011am}, consisting of a low-scale extension of the MSSM with an additional $U(1)_{\mathrm{B-L}}$ symmetry and right-handed neutrino superfields. In particular, we focus on the scalar potential induced for the $NH_uL$ flat direction, parametrized by a scalar field $\phi$ that plays the role of the inflaton and which, without loss of generality, we take to be real. This flat direction is lifted by a renormalizable term in the superpotential and by soft-SUSY breaking terms, yielding:
\begin{equation} \label{scalar_potential}
V(\phi)=\frac{1}{2}m_\phi^2\phi^2+\frac{h^2}{12}\phi^4-\frac{Ah}{6\sqrt{3}}\phi^3~. 
\end{equation}
For $A\simeq 4m_\phi$, this potential exhibits an approximate saddle point for a field value $\phi_0\simeq \sqrt{3}m_\phi/h$, such that $V'(\phi_0)\simeq V''(\phi_0)\simeq 0$, which is thus suitable for inflation. We may then define \cite{Hotchkiss:2011am}:
\begin{equation} \label{fine_tuning}
A=4m_\phi\sqrt{1-{\beta^2\over4}}
\end{equation}
and expand the potential about the generic point of inflection, yielding for $\beta\ll1$, to lowest order:
\begin{equation} \label{scalar_potential_series}
V(\phi)\simeq V_0\left(1+3\beta^2\left({\phi-\phi_0\over\phi_0}\right)+4\left({\phi-\phi_0\over\phi_0}\right)^3\right)~,
\end{equation}
where $V_0=V(\phi_0)$. This clearly shows that, for $\beta=0$, $\phi_0$ is a saddle point in the potential, with $\beta$ determining the deviations from this case, i.e. the fine-tuning of the parameters in the potential required for a sufficiently flat inflationary potential. Note that for real values of $\beta$, the potential exhibits an inflection point at $\phi_0$, whereas for imaginary values of $\beta$ it develops a local minimum at $\phi>\phi_0$, as illustrated in Fig. 1. This latter option could be suited for inflation with the field trapped in the false vacuum and then tunneling into the true minimum, as in the old inflationary picture. However, this does not lead to a graceful exit into a radiation-dominated era, so we will not consider this case in the remainder of our discussion.

\begin{figure}[htbp]
	\includegraphics[scale=1.4]{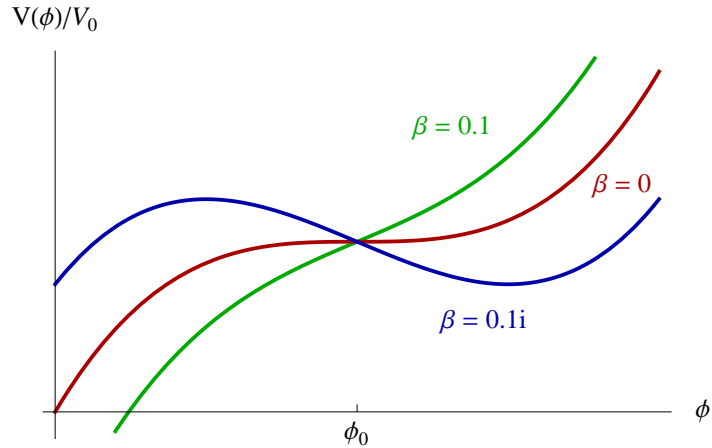} 
\caption{Normalized scalar potential for different values of the fine-tuning parameter $\beta$.}
\end{figure}
 
The inflationary dynamics, in the absence of dissipation, is determined by the slow-roll parameters, which are in this case given by:
\begin{eqnarray} \label{slow_roll_parameters}
\epsilon_\phi&=&{1\over2}m_P^2\left({V'(\phi)\over V(\phi)}\right)^2\simeq {1\over2}\left({m_P\over\phi_0}\right)^2\left(3\beta^2+12\Delta_\phi^2\right)^2~,\nonumber\\ 
\eta_\phi&=&m_P^2{V''(\phi)\over V(\phi)}\simeq 24\left({m_P\over\phi_0}\right)^2\Delta_\phi~,
\end{eqnarray}
where $m_P=2.4\times10^{18}$ GeV is the reduced Planck mass, $\Delta_\phi=(\phi-\phi_0)/\phi_0$ and we have taken $V(\phi)\simeq V_0$, which holds for $\Delta_\phi, \beta\ll1$. From these quantities we may determine the amplitude and tilt of the resulting spectrum of density perturbations, given by:
\begin{eqnarray} \label{spectrum}
\mathcal{P}_R&=&{1\over24\pi^2}{V_0/m_P^4\over \epsilon_{\phi_*}}~,\nonumber\\ 
n_s&=&1+2\eta_{\phi_*}-6\epsilon_{\phi_*}\simeq 1+48\left({m_P\over\phi_0}\right)^2\Delta_{\phi_*}~,
\end{eqnarray}
where $\phi_*$ denotes the value of the field when the relevant CMB scales left the horizon about 40-60 e-folds before the end of inflation, and we have used that $|\eta_{\phi_*}|\gg\epsilon_{\phi_*}$ for $\Delta_\phi, \beta\ll1$. These two conditions can be used to determine the constant term in the potential $V_0$ and $\phi_*$, leaving $\phi_0$ and $\beta$ as the only undetermined parameters.

The dynamics of inflation is governed by the slow-roll equation:
\begin{equation} \label{slow_roll_eq}
3H\dot\phi\simeq -V'(\phi)~, 
\end{equation}
with $H^2\simeq V(\phi)/3m_P^2$. Inflation ends in this case when the slow-roll condition $|\eta_\phi|<1$ is violated, such that $\Delta_{\phi_e}\simeq -(\phi_0/m_P)^2/24$. This allows us to compute the total number of e-folds of inflation from horizon-crossing, which is then given by:
\begin{equation} \label{e_folds}
N_e=\int_{t_*}^{t_e} Hdt\simeq -\int_{\phi_*}^{\phi_e}{3H^2\over V'(\phi)}d\phi \simeq {\arctan(1/2\xi)+\arctan\left((n_s-1)/4\xi\right)\over \xi}~,
\end{equation}
where $\xi=6\beta(m_P/\phi_0)^2$. Thus, in the limit $\beta\rightarrow0$ for any given $\phi_0$, this yields a maximum of $N_e\simeq 119$ e-folds of inflation for $n_s=0.967$ \cite{Komatsu:2010fb}, which  exceeds the observationally required range. We can invert this to determine the value of $\beta$ required for 40-60 e-folds of inflation, yielding:
\begin{equation} \label{beta_cold}
\beta\simeq (3.1-5.2)\times 10^{-3}\left({\phi_0\over m_P}\right)^2~,
\end{equation}
with smaller values of $\beta$ yielding longer periods of inflation, since the resulting potential is flatter. This illustrates the generic fine-tuning problem of inflection point models, and in this particular case the soft inflaton mass and the trilinear term in Eqs.~(\ref{scalar_potential}) and (\ref{fine_tuning}) have to compensate each other to at least one part in $10^6$ for a successful model with sub-planckian inflaton values, as can be seen by inserting Eq. (\ref{beta_cold}) in Eq. (\ref{fine_tuning}).


\section{Warm inflation dynamics}

In warm inflation, the interactions of the inflaton field with other degrees of freedom can, as discussed above, significantly affect
its dynamics and the standard inflationary observables, such as the amplitude and the tilt of the power spectrum, its non-gaussianity and
the spectrum of primordial gravitational waves. The time non-local component of these interactions leads to a transfer of energy from the inflaton field to other degrees of freedom, yielding a dissipative process which is in general described by non-equilibrium dynamics. However, in the slow-roll regime, the motion of inflaton may be sufficiently slow compared to the overall relaxation time of the system to allow for an adiabatic description of the evolution, giving to leading order an additional friction term in the scalar equation of motion:
\begin{equation}
	\ddot{\phi} + (3H + \Upsilon)\dot{\phi} + V'(\phi)=0.
 	\label{EOMf}
\end{equation}
The dissipative coefficient can be computed microscopically given a particular particle physics realization of inflation \cite{Berera:2008ar} and may in general depend on the value of the inflaton field and the properties of the multi-particle state produced by dissipation. Successful realizations of warm inflation have been constructed in a regime where the fields coupled to the inflaton acquire large masses during inflation, proportional to the inflaton value, and then decay into light degrees of freedom, which is generically known as the two-stage mechanism \cite{Berera:2002sp}. For sufficiently strong interactions the produced particles may then thermalize in less than a Hubble time, producing a quasi-thermal bath of radiation at a temperature $T$ \cite{BasteroGil:2012cm}. If the intermediate fields are sufficiently heavy compared to the temperature of the radiation, thermal corrections to the inflaton mass are Boltzmann-suppressed, thus preserving the tree-level flatness of the inflaton potential, in particular in the case of supersymmetric flat directions. Such large masses are naturally obtained for large values of the inflaton field, hence avoiding the problems of early realizations of warm inflation in a high-temperature regime \cite{Berera:1998gx, Yokoyama:1998ju}.

Supersymmetry also provides a natural framework for keeping quantum and thermal corrections to the scalar potential under control \cite{Hall:2004zr}, even though it is broken by finite temperature and energy density effects during inflation. A generic superpotential realizing the two-stage interactions is given by \cite{BasteroGil:2009ec, Moss:2008yb}:
\begin{equation}
W= W(\Phi)+g\Phi X^2 + hX Y^2~,
\label{superpot}
\end{equation}
where the scalar component of $\Phi$ is the inflaton field, its expectation value giving large masses to the bosonic and fermionic components of the intermediate superfield(s) $X$. They decay in turn into the components of the superfield $Y$, which remain light and form the radiation fluid. For $T\ll m_X$ and a broad range of couplings and field multiplicities, the dominant contribution to the dissipative coefficient corresponds to virtual excitations of the intermediate scalar fields decaying into the bosonic $Y$ components and has the form \cite{Moss:2008yb, BasteroGil:2010pb, BasteroGil:2012cm}:
\begin{equation}
	\Upsilon \approx C_{\phi}\frac{T^3}{\phi ^2}~,
	\label{dis_coeff}
\end{equation}
where $C_\phi$ is a constant that depends on the coupling $h$ and the field multiplicities in the $X$ and $Y$ sectors and which, for the purposes of our discussion, we will take as a free parameter of the model. Note that renormalizable superpotentials of the form in Eq.~(\ref{superpot}) are ubiquitous in supersymmetric models, such as for example the NMSSM, where the additional singlet could play the role of the inflaton and dissipate its energy into (s)quarks and (s)leptons through the Higgs portal, e.g. $W=g\Phi H_uH_d+hQH_uU+\ldots$ Notice, however, that a much larger number of fields is required with the dissipation coefficient in Eq. (\ref{dis_coeff}) to achieve a sufficient number of e-folds of inflation than in the MSSM. More generically, one expects fields coupled to supersymmetric flat directions to acquire large masses during inflation (see e.g. \cite{Allahverdi:2007zz}), while fields that do not couple directly to flat directions should remain light. Such a superpotential also arises in D-brane constructions, where dissipative effects have been shown to play an important role in overcoming the associated eta-problem \cite{warm_brane}.

The thermalized radiation fluid has an energy density
\begin{equation}
	\rho_r \simeq \frac{\pi^2}{30}g_*T^4~,
	\label{rad}
\end{equation}
where $g_*$ is the effective number of light degrees of freedom, and is sourced by the dissipative motion of the inflaton field, yielding
\begin{equation}
	\dot{\rho}_r + 3H(\rho_r + p_r)=\Upsilon\dot{\phi}^2,
	\label{EOMr}
\end{equation}
where $p_r$ is the pressure associated with the radiation fluid. In warm inflation, this fluid is not redshifted away during inflation, due to the additional dissipative source term \cite{wi, wi2}. The radiation energy density needs, however, to be subdominant to achieve a period of accelerated expansion, i.e. $\rho_r\ll \rho_\phi$, where $\rho_\phi=\dot{\phi}^2/2 + V(\phi)$. However, the associated temperature may be  larger than the expansion rate, $T> H$, which allows one to neglect the effects of expansion in computing the dissipation coefficient and modifies the evolution of inflaton fluctuations. Otherwise, when $T< H$, dissipative effects can be neglected and the standard cold inflation scenario is recovered.

Slow-roll inflation, whether cold or warm, requires an overdamped evolution of the inflaton field. In warm inflation this can be achieved due to the friction term $\Upsilon$ in addition to Hubble damping. Once the field $\phi$ is in the slow-roll regime, the evolution of the radiation fluid is also generically damped, and the equations of motion reduce to
\begin{align}
3 H ( 1 + Q ) \dot \phi &\simeq  -V'(\phi)    \,,\label{eominfsl} \\
4 \rho_R  &\simeq 3 Q\dot \phi^2\,, \label{eomradsl}
\end{align}
where we have introduced the dissipative ratio $Q=\Upsilon/(3H)$, which, depending on the particular
model, may increase or decrease during inflation. The slow-roll conditions are now given by $\epsilon_\phi, \eta_\phi\ll 1+Q$, alleviating the need  for a very flat potential, and additionally one requires that \cite{Moss:2008yb}:
\begin{eqnarray}
\beta_\Upsilon &=& m_P^2 \left ( \frac{\Upsilon_\phi V_\phi }
     {\Upsilon V}\right)\ll 1+Q \,, \label{beta} \\
\delta &=& \frac{T V_{T\phi}}{V_\phi} < 1  \label{delta}\,,
\end{eqnarray}
where $\beta_{\Upsilon}$ measures the variation of the dissipation coefficient with respect to the inflaton field and the last condition ensures that thermal corrections to the potential are small, which is the case in the regime where Eq.~(\ref{dis_coeff}) is valid.


\section{Warm inflation near an inflection point}

As previously discussed, the additional friction term in Eq. (\ref{EOMf}) alleviates the flatness of the potential required in order to achieve a sufficient amount of inflation. In the context of inflection point inflation, we have seen that the $\beta$ parameter determines the shape of the potential in the vicinity of the inflection point, measuring the fine-tuning of the underlying parameters. Therefore, we expect that a warm realization of these models can naturally reduce the amount of fine-tuning required.

We will use the Eq. (\ref{scalar_potential}) as a working example of a potential with an inflection point to analyze the generic dynamics of warm inflation in this context, although this does not correspond to a concrete realization of warm inflation in the MSSM. Writting Eq. (\ref{scalar_potential}) in the form of Eq. (\ref{scalar_potential_series}), the dynamics is described by six independent parameters, in particular the value of the field at the inflection point $\phi_0$, the corresponding height of the potential $V_0$, the fine-tuning parameter $\beta$, the value of the field at horizon-crossing $\phi_*$, the dissipative constant $C_\phi$ and the effective number of light degrees of freedom $g_*$. We can use the WMAP 7-year results giving a power spectrum with an amplitude $\mathcal{P}_{\mathcal{R}}=(2.43\pm0.11)\times 10^{-9}$ and a spectral index $n_s =0.967\pm0.014$ \cite{Komatsu:2010fb} to determine $V_0$ and $\phi_*$. As mentioned above, for $T>H$ the dominant contribution to the spectrum of primordial perturbations are the thermal fluctuations in the coupled inflaton-radiation system, and the resulting amplitude of the power spectrum in the slow-roll regime is given by \cite{wi, Berera:1999ws, Hall:2003zp}:
\begin{equation}
\mathcal{P}^{1/2}_\mathcal{R}\simeq\left(\frac{H}{2\pi}\right)\left(\frac{3H^2}{V'(\phi)}\right)(1+Q)^{5/4}\left(\frac{T}{H}\right)^{1/2}~,
	\label{amplitude}
\end{equation}
where all quantities are implicitly evaluated at horizon-crossing. In order to solve this equation, it is useful to write it in a more convenient way. Using the slow-roll equations (\ref{eominfsl}) and (\ref{eomradsl}), one obtains
\begin{equation}
	Q_*(1+Q_*)^{13/2} \simeq \mathcal{P_{R}}\left(\frac{C_\phi}{3}\right)\left(\frac{\pi C_\phi}{2C_R}\right)^2(2\epsilon_{\phi_*})^3\left(\frac{m_P}{\phi_*}\right)^6~,
	\label{qphi}
\end{equation}
where $C_R=g_*\pi^2/30$. Eq. (\ref{qphi}) and the expression for the spectral index \cite{BasteroGil:2009ec}
\begin{equation}
	(1+Q_*)(1+7Q_*)(n_s-1)+(2+9Q_*)\epsilon_{\phi_*} + 3Q_*\eta_{\phi_*} + (1+9Q_*)\beta_{\Upsilon_*}\simeq 0
	\label{index}
\end{equation}
form a coupled system of equations for $Q_*$ and $\phi_*$ that needs to be solved numerically for given values of $\phi_0$, $\beta$, $C_\phi$ and $g_*$. Once the system is solved, we can obtain the value of $V_0$ using Eq. (\ref{amplitude}):
\begin{equation}
	V_{0} \simeq \left(\frac{C_R}{C_\phi}\right)\frac{144\pi^2\mathcal{P_R}\phi^2_*m^2_P}{\sqrt{1+Q_*}\left(1+3\beta^2\left({\phi_*-\phi_0\over\phi_0}\right)+4\left({\phi_*-\phi_0\over\phi_0}\right)^3\right)}~.
	\label{vnot}
\end{equation}
The system has in general three possible solutions satisfying the observational constraints, and we have consistently chosen the one that maximizes the difference $\phi_* - \phi_0$, since as we discuss below this minimizes the amount of dissipation required for a sufficiently long period of inflation. To simplify the numerical procedure, one can use the approximate solutions in the strong and weak dissipative regimes, $Q_*\ll1$ and $Q_*\gg 1$, respectively, where the equations decouple, to find the initial guess required to calculate numerically the full solution to the coupled system of equations. In the intermediate regime, $Q_*\sim 1$, it is sufficient to use an initial guess in this range. 


\section{Numerical results}

\begin{figure}[htbp]
	\includegraphics[angle=-90, scale=0.17]{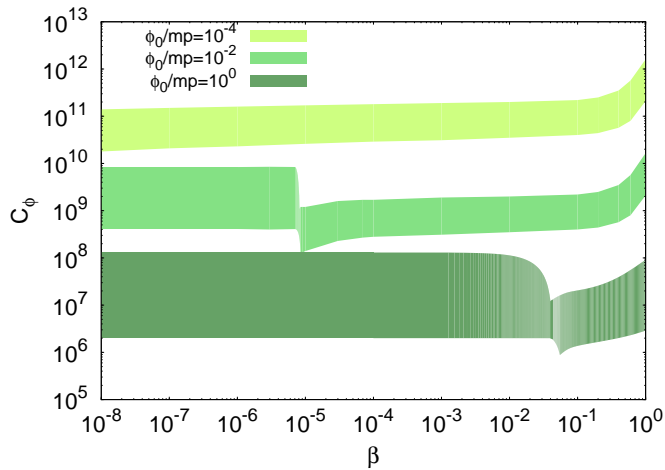} 
\caption{Values of $C_\phi$ and $\beta$ required to obtain $N_e\in[40,60]$ for $g_*=100$ and different values of $\phi_0/m_P$.}
\label{variop}
\end{figure}

Having determined $V_0$ and $\phi_*$ from the observational constraints, we may now study the evolution of the coupled inflaton-radiation system as a function of the remaining parameters, $C_\phi$, $\beta$, $\phi_0$ and $g_*$. For concreteness, we first fix the number of relativistic degrees of freedom  $g_*=100$, corresponding to the order of magnitude of the number of MSSM scalar fields, although we study the effect of varying this parameter at the end of this section. Our main goal is to determine which is the lowest value of $C_\phi$ required for a sufficiently long period of inflation as a function of the fine-tuning parameter $\beta$ and for different values of $\phi_0$. The number of e-folds of warm inflation can be computed by including the effects of dissipation in Eq.~(\ref{e_folds}):
\begin{equation} \label{e_folds_warm}
	N_e \simeq -\int_{\phi_*}^{\phi_e}{\frac{3H^2(1+Q)}{V'(\phi)}d\phi}.
\end{equation}
However, due to the $T$- and $\phi$-dependent dissipative ratio $Q$, this integral cannot be solved analytically as in the cold inflation case. Besides, the value of the field at the end of inflation cannot be calculated {\it a priori}. Hence, the equations of motion for both the inflaton and the radiation fluid have to be integrated numerically. In most areas of the parameter space, the inflaton field is always in the slow-roll regime and therefore we may integrate Eq. (\ref{eominfsl}). However, in some regions of the parameter space the radiation fluid is not slow-rolling, in which case we integrate the full equation (\ref{EOMr}). The consistency of our analysis is determined by three main conditions:
\begin{itemize}
	\item $\epsilon_H=-\frac{\dot{H}}{H} < 1$, which is required for accelerated expansion and generalizes the slow-roll conditions;
	\item $m_{X}\gg T$, for which the form of the dissipative coefficient in Eq. (\ref{dis_coeff}) is valid.
	\item $T>H$, which allows one to neglect the effects of expansion in computing the dissipation coefficient and generically defines the regime where inflation is warm.
\end{itemize}
These conditions need to hold for $40-60$ e-folds of inflation in order to solve the horizon and flatness problems, and in Fig. \ref{variop} we show the regions in the plane $C_{\phi}-\beta$ where this is obtained for different values of the inflection point $\phi_0$.

As one can easily see in this figure, lower values of $\phi_0$ require more dissipation in order to obtain the same number of e-folds, which is related to the associated increase in the slow-roll parameters in Eq.~(\ref{slow_roll_parameters}), due to a steeper shape of the potential. In addition, we find two distinct regions of parameter space in Fig. \ref{variop}, corresponding to small and large values of the fine-tuning parameter $\beta$. The separation between these regions depends on the value of $\phi_0$, with the small-$\beta$ region moving to lower values of $\beta$ for smaller $\phi_0$. 

\begin{figure}[htbp]
\subfiglabelskip=5pt
	\centering \subfigure[$(\phi-\phi_0)/\phi_0$]  { \centering
  \includegraphics[scale=0.1065, angle=-90]{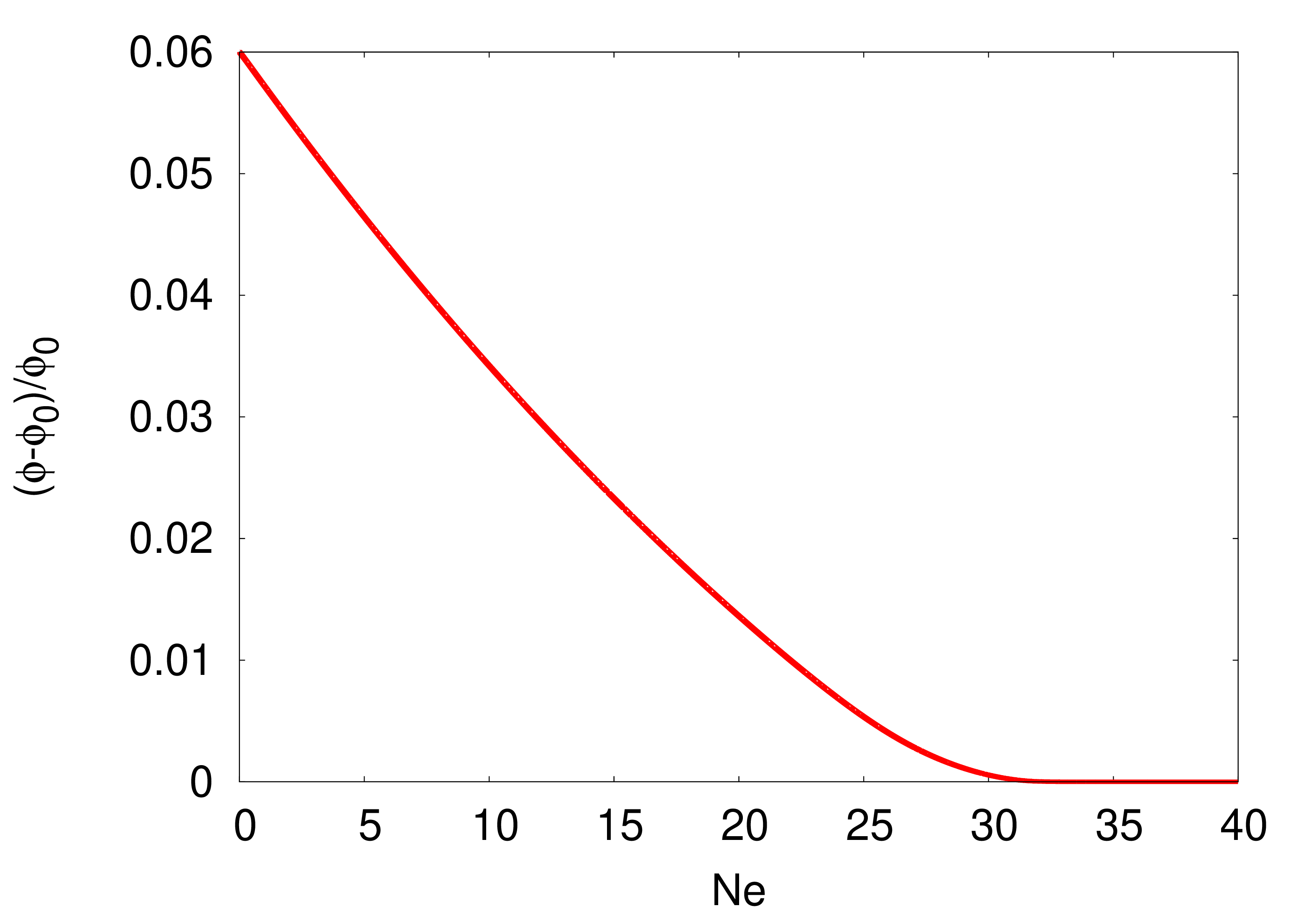}
\label{p1e7}
}\subfigure[$\epsilon_H$ and $\rho_r/\rho_\phi$] { \centering \includegraphics[scale=0.1065,
    angle=-90]{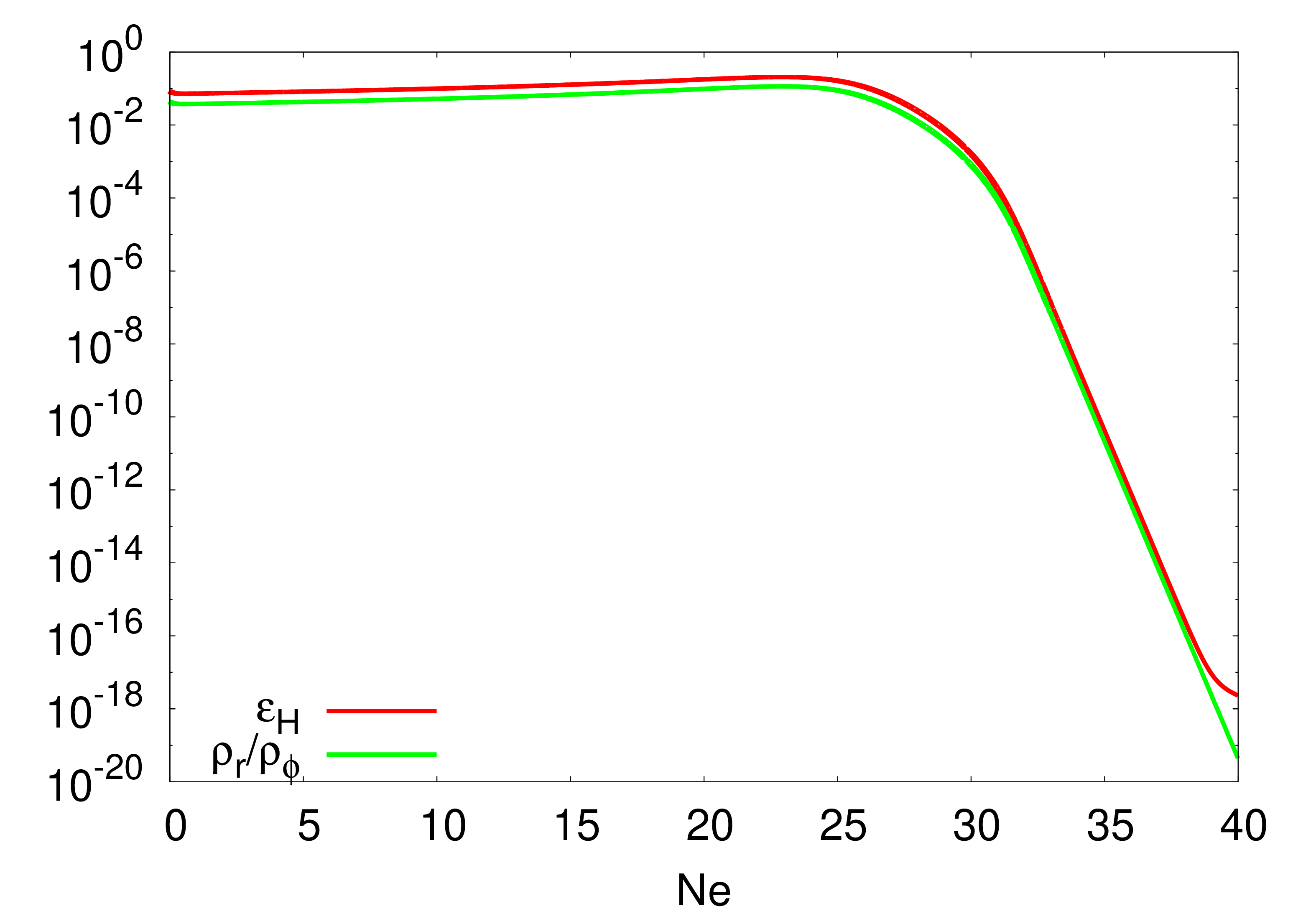}
\label{e1e7}
}
\subfigure[$T/H$] { \centering \includegraphics[scale=0.1065,
    angle=-90]{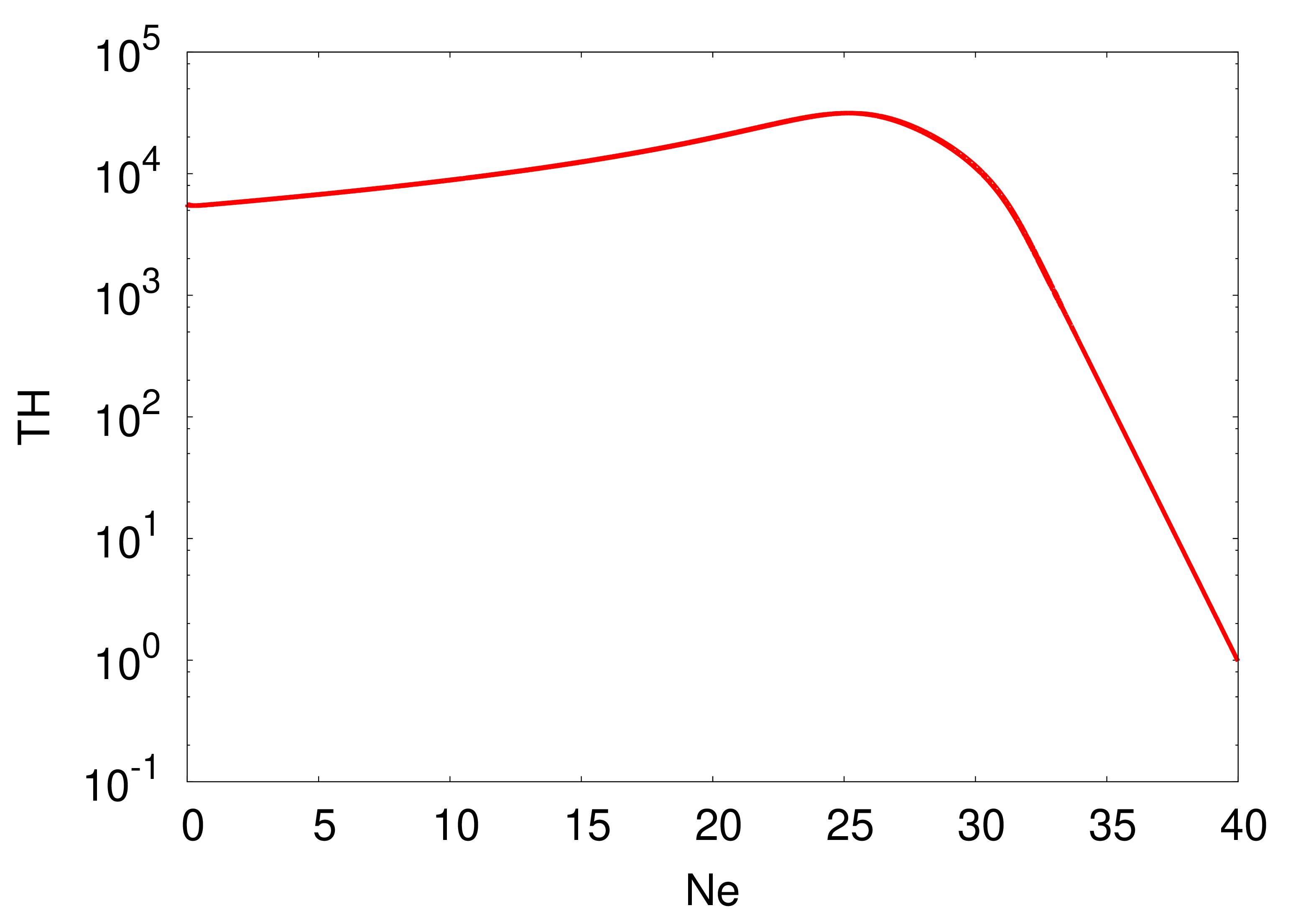}
\label{t1e7}
}
\caption{Evolution with the number of e-folds of $(\phi-\phi_0)/\phi_0$, $\epsilon_H$, $\rho_r/\rho_\phi$ and $T/H$ for $\phi/m_P=10^{-2}$, $g_*=100$ and $\beta=10^{-7}$ when inflation lasts for $40$ e-folds.}
\label{1e7}
\end{figure}

In the small-$\beta$ region, the potential is extremely flat and intuitively one would expect less friction to be required for a given period of accelerated expansion. However, Fig. \ref{variop} clearly shows that the required value of $C_\phi$ becomes constant for low values of $\beta$, which suggests taking a closer look at the physical mechanism behind dissipation. Since it is the motion of the inflaton field that produces light particles in a quasi-thermal bath, the amount of radiation produced depends on how fast the inflaton is rolling, as can be explicitly seen in Eq.~(\ref{EOMr}). If the potential is too flat, the inflaton will roll too slowly, which suppresses the amount of radiation produced and consequently decreases the temperature of the thermal bath. In fact, it is the condition $T>H$ that determines the end of warm inflation in this region of parameter space, as one can see in Fig. \ref{1e7}, where we plot the evolution of the relevant quantities in this regime. This also explains why the initial condition farther away from the inflection point yields the lowest value of $C_\phi$, since an initially steeper potential can more easily produce a radiation bath with $T>H$.

In Fig. \ref{p1e7}, one can see that the inflaton starts above the inflection point and ends close to the latter, with the temperature dropping below the Hubble rate after 40 e-folds of inflation. Notice, however, that inflation does not necessarily end at this point, since $\epsilon_H<1$ and decreasing, but our analysis is no longer consistent at this stage since de Sitter effects may modify the dissipation coefficient. It may, in fact, be possible for an additional period of cold inflation to follow, thus decreasing the amount of dissipation required to achieve the desired number of e-folds, although a detailed analysis of this possibility is beyond the scope of this work.  Finally, in Fig. \ref{1e7} we see that $\epsilon_H$ follows closely the evolution of the radiation energy density, which in this case is becoming more and more sub-leading compared to the inflaton field, thus requiring an additional reheating stage.

\begin{figure}[htbp]
\subfiglabelskip=5pt
	\centering \subfigure[$(\phi-\phi_0)/\phi_0$]  { \centering
  \includegraphics[scale=0.1065, angle=-90]{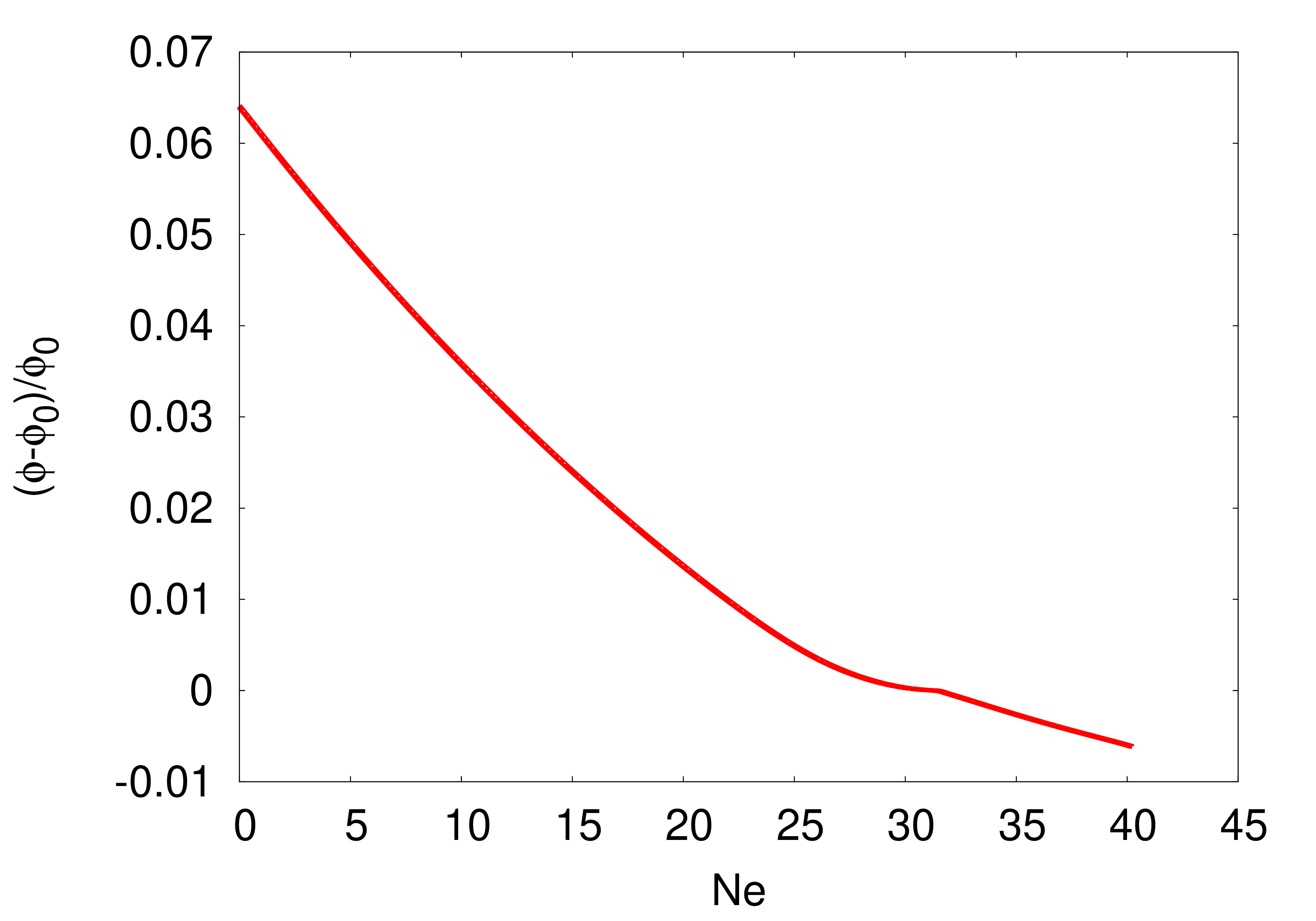}
\label{p1e2}
}\subfigure[$\epsilon_H$ and $\rho_r/\rho_\phi$] { \centering \includegraphics[scale=0.1065,
    angle=-90]{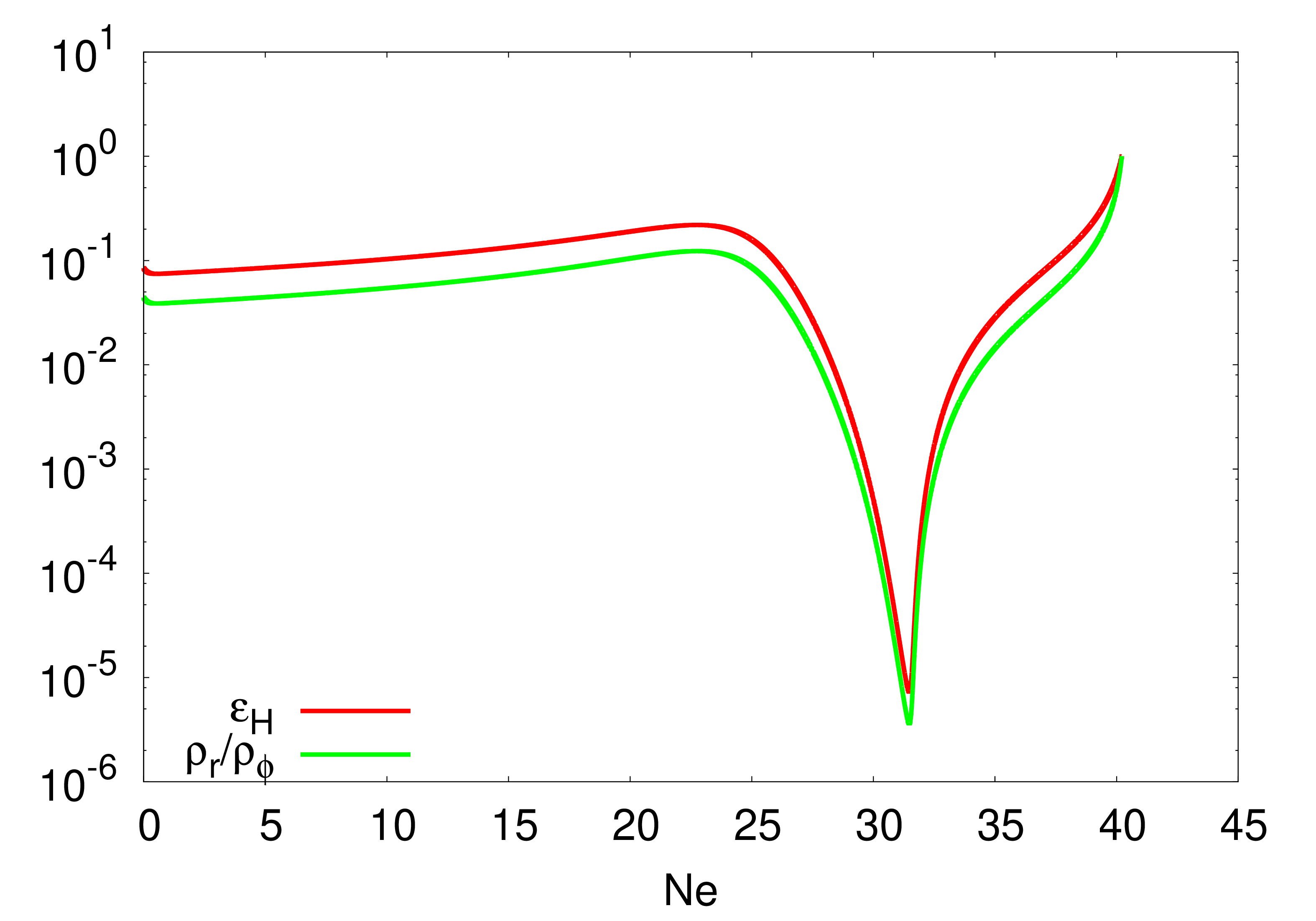}
\label{e1e2}
}
\subfigure[$T/H$] { \centering \includegraphics[scale=0.1065,
    angle=-90]{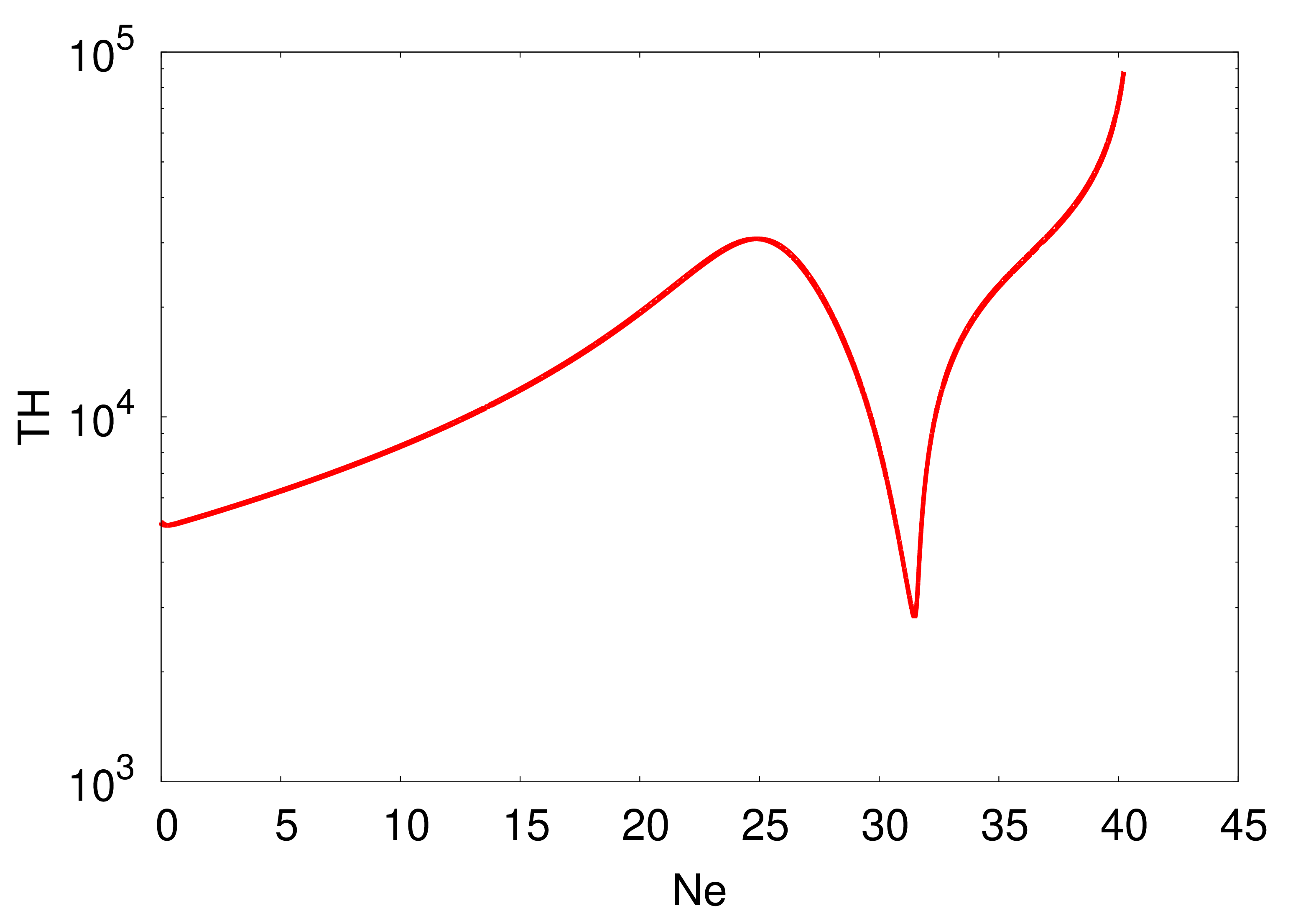}
\label{t1e2}
}
\caption{Evolution with the number of e-folds of $(\phi-\phi_0)/\phi_0$, $\epsilon_H$, $\rho_r/\rho_\phi$ and $T/H$ for $\phi/m_P=10^{-2}$, $g_*=100$ and $\beta=10^{-2}$ when inflation lasts for $40$ e-folds.}
\label{1e2}
\end{figure}

In the large-$\beta$ region the potential is steeper, therefore the production of radiation is enhanced and $T>H$ is no longer the dominant constraint. In fact, in this regime radiation tends to be overproduced and dominate the energy density, thus allowing for a graceful exit from  inflation, as shown in Fig. \ref{1e2} where we plot the evolution with the number of e-folds of the relevant quantities in the large-$\beta$ region.

In Fig. \ref{p1e2}, one can see that the inflaton field starts away from the inflection point, remains close to it for a few e-folds but that, due to the slope of the potential, inflation ends beyond the point of inflection, in contrast with the small-$\beta$ behavior. Notice that $\rho_r/\rho_\phi$ decreases sharply when the field slows down close to the inflection point, in agreement with the discussion above, but then increases as the field moves to lower values and  eventually ends inflation with a smooth exit into a radiation-dominated era. In Fig. \ref{t1e2} it is also clear that $T>H$ for the whole duration of inflation.

Although we have not plotted the condition $m_{X}\gg T$ in Figs.  \ref{1e7} and \ref{1e2}, we have checked that it is satisfied in all  the parameter space shown, for couplings $g\sim1$. On the other hand,  we may consider more general potentials, associated with different SUSY breaking effects, yielding a different value for the numerical coefficient of the slow-roll parameter $\eta_\phi$ in Eq.~(\ref{slow_roll_parameters}). We then find that, for lower values of this coefficient, the condition $m_X\gg T$ is more stringent than $T>H$.  However, the amount  of dissipation required does not change significantly even for an order of magnitude change in this coefficient, so we do not explore  this possibility in more detail.

\begin{figure}[h]
	\includegraphics[angle=-90, scale=0.17]{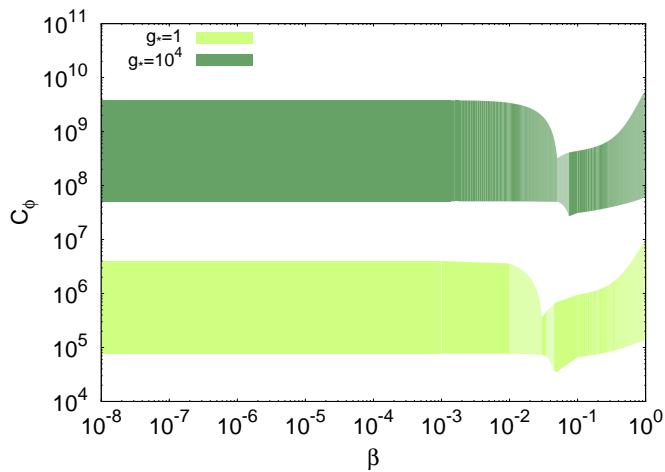} 
\caption{Values of $C_\phi$ and $\beta$ required to obtain $N_e\in[40,60]$ for $\phi_0/m_P=1$ and $g_*=1,10^4$.}
\label{variog}
\end{figure}

Finally, we analyze the effect of the number of relativistic degrees of freedom on the amount of dissipation required for successful inflation. In Fig. \ref{variog} we show the $C_\phi-\beta$ region where 40-60 e-folds of inflation are obtained with different values of $g_*$. 

In Fig. \ref{variog}, it can be observed that the required value of $C_\phi$ decreases for smaller $g_*$. In order to understand this behavior, we compute the explicit dependence of the dissipation coefficient  on $g_*$ by substituting Eq. (\ref{rad}) into Eq. (\ref{dis_coeff}):
\begin{equation}
	\Upsilon = \frac{30^{3/4}C_\phi}{\pi^{3/2} g_*^{3/4}}\frac{\rho^{3/4}_r}{\phi^2}.
\end{equation}
Hence, the relevant quantity is an effective dissipation constant:
\begin{equation}
	\tilde{C}_{\phi}=\frac{C_\phi}{g^{3/4}_*}
\end{equation}
that remains constant in Fig. \ref{variog} for the different values of $g_*$, which is also the case for smaller (sub-planckian) values of the inflection point.


\section{Summary and future prospects}

In this work we have analyzed the dynamics of inflation close to an inflection point in the scalar potential taking into account dissipative effects resulting from the coupling of the inflaton field to other degrees of freedom. We have focused on supersymmetric models, where the plethora of available flat directions may be lifted by competing SUSY-breaking effects, producing inflection and even saddle points in the potential, although at the expense of fine-tuning {\it a priori} unrelated parameters. Moreover, supersymmetry provides a natural framework for warm inflation, helping to protect the flatness of the potential against both thermal and radiative corrections, in particular in the regime where the fields coupled to the inflaton acquire large masses and their effects are Boltzmann-suppressed, nevertheless allowing for the dissipative production of virtual excitations that may decay into light fields in a quasi-thermal bath. For concreteness, we have focused on the $NH_uL$ flat direction in a low scale extension of the MSSM, although the resulting scalar potential has a sufficiently generic form and our main results should apply to other realizations of inflection point inflation. 

Our numerical simulations of the dissipative dynamics of inflation in this model have lead us to two main conclusions. Firstly, if dissipative effects are sufficiently strong, a sufficiently long period of inflation may occur independently of the fine-tuning of the parameters in the potential, which was expected since the additional friction alleviates the need for a very flat potential. Secondly, and more surprisingly, the required amount of dissipation does not decrease arbitrarily for flatter potentials, given that if the scalar potential is too flat and the inflaton evolves too slowly, it becomes more difficult to sustain a radiation bath with a temperature above the Hubble rate, which is required for consistency of our analysis. This results in a field-dependent critical value of the fine-tuning parameter $\beta$ below which the required dissipation parameter $C_\phi$ becomes constant. Above this value, the potential is sufficiently steep to ensure that $T>H$ throughout inflation, with steeper potentials requiring larger values of the dissipation parameter.

The value of $C_\phi$ depends on the coupling between the intermediate fields and the light degrees of freedom, as well as on the multiplicities of both heavy and light fields. The minimum value of $C_\phi\gtrsim 10^6$ obtained for $g_*=100$ is of the same order as that obtained for other forms of the inflaton potential, such as monomial or hybrid models \cite{BasteroGil:2009ec}, which implies large couplings and field multiplicities, so one may ask whether there is any gain from the model building perspective in trading a large fine-tuning in the parameters of the potential for large couplings and a large number of fields. On one hand, fine-tuning makes inflation less generic, since it isolates a small region of the available parameter space, whereas inflation should provide an explanation for the otherwise finely-tuned conditions in the early universe. On the other hand, a large number of degrees of freedom during inflation points towards more complicated beyond the Standard Model scenarios, e.g.~with fields in large representations, which may be realized in generic GUT constructions or D-brane models \cite{warm_brane}. As discussed earlier in this work, strong dissipative effects may have other interesting consequences in the dynamics of the early universe and, moreover, we have seen that the minimum value of $C_\phi$ may be substantially reduced in models with a smaller number of relativistic degrees of freedom, where the temperature of the radiation bath is consequently larger, pointing towards constructions with several fields coupled directly to the inflaton but with few distinct decay channels. 

One should point out that we have focused on a particular form of the dissipation coefficient, corresponding to the decay of virtual excitations of the heavy fields coupled to the rolling inflaton. In \cite{BasteroGil:2012cm}, it was pointed out that this is not the only possibility, as the excitation of real modes becomes the dominant contribution to dissipative effects for $m_X/T\lesssim 10$, yielding a much larger dissipation coefficient than virtual excitations for $\mathcal{O}(1)$ couplings and field multiplicities, nevertheless suppressing quantum and thermal corrections to the scalar potential. While the dynamics of warm inflation in this regime remains unexplored, it suggests a much more promising avenue from the model building point of view, and we expect our main qualitative results to apply in this case as well, since they depend more on the presence of significant dissipative effects than on the specific form of the dissipation coefficient. 

Whereas we intend to explore this regime in more detail in the future, our results show that dissipative effects are an interesting and well-motivated alternative to fine-tuning in inflationary models and we hope to motivate further investigation of this topic in other inflationary scenarios.


\begin{acknowledgements}
We would like to thank Mar Bastero-Gil, Arjun Berera and Anupam Mazumdar for their useful comments and suggestions. R.C. is partially supported by MICINN (FIS2010-17395) and ``Junta de Andaluc\'ia" (FQM101) and would like to acknowledge the hospitality of the Particle Theory Group at the University of Edinburgh during the completion of this work. The work of J.G.R. was partially supported by the STFC Particle Theory Group grant at the University of Edinburgh (United Kingdom), the FCT grant PTDC/FIS/116625/2010 (Portugal) and the Marie Curie action NRHEPÐ295189- FP7-PEOPLE-2011-IRSES.

\end{acknowledgements}


\end{document}